\documentclass[extra,referee]{gji}
\usepackage{url} 

\usepackage{timet}
\usepackage{amsmath, enumitem}
\usepackage{amsfonts}
\usepackage{upgreek}
\usepackage{graphicx}
\graphicspath{{Figures/}} 
\usepackage{setspace}
\usepackage{multirow}
\usepackage[T1]{fontenc}

\begin{document}
	\begin{titlepage}
		\begin{center}
			\vspace*{-2.0cm} 
			
			
			\vspace{5cm}
			
			\huge
			\textbf{Rapid Bayesian Seismic Tomography using Graph Mixture Density Networks}
			
			\vspace{4.0cm}
			\LARGE
			Xin Zhang$^1$, Yan Wang$^{2}$, Haijiang Zhang$^{2}$
			
			\vspace{1cm}
			\Large
			$^1$ School of Engineering and Technology, China University of Geosciences, Beijing, China \\
			$^2$ Laboratory of Seismology and Physics of Earth’s Interior, School of Earth and Space Sciences, University of Science and Technology of China, Hefei, China \\
			
			\vspace{1cm}
			\Large
			E-mail: \textit{xzhang@cugb.edu.cn}
			
			\vfill
			\vfill
		\end{center}
		
	\end{titlepage}

\begin{abstract}
Seismic tomography is a methodology to image subsurface properties of the Earth. In order to better interpret the resulting images, it is important to assess uncertainty in the results. Mixture density networks (MDNs) provide an efficient way to estimate Bayesian posterior probability density functions (pdfs) that describe the uncertainty of tomographic images. However, the method can only be applied in cases where the number of data is fixed, and consequently a large number of practical applications that have variable data sizes cannot be solved. To resolve this issue, we introduce graph neural networks (GNNs) to solve seismic tomographic problems. Graphs are data structure which provides flexible representation of complex, variable systems. GNNs are neural networks that manipulates graph data, and can be combined with MDNs (called graph MDNs) to provide efficient estimates of posterior pdfs for graph data. In this study we apply graph MDNs to seismic tomography by representing travel time data with a graph. We demonstrate the method using both synthetic and real data, and compare the results with those obtained using Markov chain Monte Carlo (McMC). The results show that graph MDNs can provide comparable posterior pdfs to those obtained using McMC at significantly lower cost. We thus conclude that graph MDNs can be used in a range of practical applications that require many similar seismic tomographic problems with different number of data to be solved.  

\end{abstract}

\section{Introduction}
In a variety of academic and practical applications, geoscientists often need to find answers to scientific questions that concern the Earth's subsurface. A common approach is to image the subsurface properties using data recorded at or above the Earth's surface, and to interpret the results to address questions of interest. Seismic tomography is one such method which has been used widely to generate those images. In order to obtain robust and well-justified answers, it is important to assess uncertainties in tomographic results. 

Many modern applications require tomographic problems to be solved efficiently. For example, in a  monitoring scenario that aims to detect dynamic changes of the subsurface, it is often required to image the subsurface properties in near real time \citep{duputel2009real, cao2020near, hanafy2021near}. In addition, the advance of acquisition equipment has led to a large amount of data being recorded continuously, which need to be processed and inverted in an efficient way \citep{arrowsmith2022big}.

Tomographic problems are traditionally solved using optimization methods in which one seeks an optimal set of model parameters by minimizing the difference between observed data and model predicted data \citep{aki1976determination, dziewonski1987global, iyer1993seismic}. Because of nonuniqueness and nonlinearity of tomographic problems, some form of regularization is often used to solve the system \citep{aki1976determination, tarantola2005inverse, aster2018parameter}. However, the regularization is often chosen by ad hoc means (often by trial and error), which makes it difficult to perform the procedure automatically and efficiently, and can suppress valuable information in the data \citep{zhdanov2002geophysical}. In addition, optimization methods cannot provide accurate uncertainty estimates for nonlinear problems because the methods often use linearized physics to assess uncertainty of solutions. To resolve these issues, partially or fully nonlinear methods based on Bayesian formulation of inverse problems have been introduced to seismic tomography. In such methods one finds a so-called posterior probability density function (pdf) which describes the distribution of possible models, by combining prior information with new information contained in the data. These include Markov chain Monte Carlo (McMC) methods \citep{mosegaard1995monte, malinverno2000monte, bodin2009seismic, piana2015local, zhang20183} and variational inference \citep{nawaz2019rapid, zhang2020seismic, zhang20233}.

McMC is a class of methods which generate a set (or chain) of successive samples from the posterior pdf \citep{metropolis1949monte,hastings1970monte,brooks2011handbook}; those samples can thereafter be used to infer useful statistics of that pdf. The methods are quite general from a theoretical point of view, and have been applied to solve a wide range of seismic tomographic problems \citep{bodin2009seismic, galetti2015uncertainty, piana2015local, fichtner2018hamiltonian, burdick2017velocity, zhang20183, zhang2020imaging}. However, such solutions are obtained at significant expense, typically requiring days to weeks of computer run time, and hence cannot be applied in scenarios where rapid solutions are required \citep{duputel2009real, kaufl2016solving, arrowsmith2022big}.

Variational inference solves Bayesian inference problems using optimization: within a predefined (often simplified) family of pdfs, the method seeks an optimal approximation to the posterior distribution by minimizing the difference between the approximating pdf and the posterior pdf \citep{bishop2006pattern, blei2017variational}. The method has been demonstrated to be more efficient than McMC sampling methods for certain types of problems \citep{bishop2006pattern, blei2017variational}. In geophysics, variational inference has been applied to petrophysical inversion \citep{nawaz2019rapid, nawaz2020variational}, seismic tomography \citep{zhang2020seismic, zhao2022bayesian, agata2023bayesian, zhang2024vip}, full waveform inversion \citep{zhang2020variational, urozayev2022reduced, zhang20233, izzatullah2024physics}, geostatistical inversion \citep{liu2024geostatistical} and seismic imaging \citep{siahkoohi2020faster}. Nevertheless, the method still requires a large amount of computational time, and therefore cannot provide sufficiently rapid solutions for real time monitoring, nor for cases where many similar inversions are required.   

Neural network-based methods provide an efficient alternative for certain classes of problems that need to be solved many times with new data of similar type. An initial set of model parameters is generated from prior information, and data are simulated from these model parameters using known physical relationships. Neural networks are flexible mappings which can be trained to emulate the inverse mapping from data to parameter space by fitting the set of example pairs of data and model parameters \cite[called training data set;][]{bishop2006pattern}. After training, the networks can be used to provide efficient estimate of model parameter values for any new, measured data. The method can therefore be used in applications that require solutions of many inverse problems within the class of problems represented by training data. 

Neural network-based methods have been used to solve a variety of geophysical inverse problems, including seismic tomography \citep{roth1994neural, moya2010inversion, araya2018deep}, electromagnetic inversion \citep{puzyrev2019deep}, gravity inversion \citep{huang2021deep}, surface wave dispersion inversion \citep{hu2020using} and full waveform inversion \citep{yang2019deep}. In addition, neural networks can also be used within the framework of inversion to improve accuracy of results. For example, the method has been used to create new regularization for traditional inversion \citep{zhang2022regularized}, to extrapolate low frequency data from high frequency signals \citep{ovcharenko2019deep}, to reparameterize seismic velocity models \citep{he2021reparameterized, sun2023implicit}, and to encode geological information into prior information \citep{laloy2017inversion, mosser2020stochastic}. Nevertheless, these studies did not provide estimates of uncertainty as only a single set of model parameters is obtained for each data set.

To enable uncertainty estimate in neural network-based inversions, \cite{devilee1999efficient} proposed a form of probabilistic neural networks which provide discretized estimates of posterior pdfs, and applied the method to surface wave dispersion inversion. In an alternative formulation, mixture density networks (MDNs) output a probability distribution which is represented by a sum of analytic pdfs called kernels, such as Gaussian distributions, and can be trained to provide estimate of posterior distributions for given data \citep{bishop2006pattern}. MDNs have been applied to many geophysical inverse problems, such as surface wave dispersion inversion \citep{meier2007aglobal, meier2007bglobal, earp2020probabilisticGrane}, travel time tomography \citep{earp2020probabilistic}, petrophysical inversion \citep{shahraeeni2011fast, shahraeeni2012fast}, earthquake source parameter inversion \citep{kaufl2014framework, kaufl2015robust}, pore pressure prediction \citep{karmakar2019short}, Earth's radial seismic structure inversion \citep{de2013bayesian} and acoustic-articulatory inversion \citep{richmond2007trajectory}. In addition, \cite{ardizzone2018analyzing} proposed another method which uses invertible neural networks (INNs) to predict posterior distributions. INNs are a class of networks that provide bijective (two-way) mappings between inputs (models) and outputs (data), and can be trained to provide estimates of posterior distributions by adding latent variables in the data side. In geophysics the method has been applied to surface wave dispersion inversion \citep{zhang2021bayesianb}, travel time tomography \citep{zhang2021bayesianb, sun2024invertible} and airborne electromagnetic inversion \citep{wu2023fast}. However, in above studies the trained MDNs or INNs require a fixed size of data, and consequently the methods cannot be applied in cases where the number of data varies, which is common in practice because of noise or malfunction of instruments.

To extend neural network-based inversion to applications with variable sizes of data, we introduce graph neural networks (GNNs) to solve geophysical inverse problems. Graphs are a type of data structure which provides flexible representation to complex, variable systems. GNNs are neural networks that manipulate graphs. By combining GNNs and MDNs, one can create a graph mixture density network which can be trained to provide estimates of posterior pdfs for graph data \citep{errica2021graph}. In this study we apply graph MDNs to seismic travel time tomography by using graphs to represent travel time data that often have variable sizes in practice.

In the next section we describe the basic idea of MDNs and graph MDNs. We then apply the method to a synthetic 2D travel time tomographic example with variable data distributions, and compare the results with those obtained using McMC. The results show that graph MDNs can provide comparable solutions to those obtained using McMC at significantly reduced computational cost. In section 4, we train a graph MDN to predict posterior distributions of group velocity at different periods in South England, which further demonstrates that graph MDNs can provide rapid, accurate approximations of posterior pdfs for problems that have variable number of data. We thus conclude that graph MDNs provide an important tool to solve many similar tomographic problems efficiently.  

\section{Methods}
\subsection{Bayesian inference}
Bayesian inference is the process of constructing a posterior probability density function (pdf) $p(\mathbf{m}|\mathbf{d}_{obs})$ of model parameters $\mathbf{m}$ given the observed data $\mathbf{d}_{obs}$. This is achieved by updating a \textit{prior} pdf 
$p(\mathbf{m})$ with new information contained in the data $\mathbf{d}_{obs}$ using Bayes' theorem,
\begin{equation}
    p(\mathbf{m}|\mathbf{d}_{obs}) = \frac{p(\mathbf{d}_{obs}|\mathbf{m})p(\mathbf{m})}{p(\mathbf{d}_{obs})}
\label{eq:Bayes}
\end{equation}
where $p(\mathbf{d}_{obs}|\mathbf{m})$ is the \textit{likelihood} which describes the probability of observing data $\mathbf{d}_{obs}$ if model $\mathbf{m}$ was true, and $p(\mathbf{d}_{obs})$ is a normalization factor called the \textit{evidence}. 
The likelihood function is often assumed to follow a Gaussian distribution around the data predicted synthetically (using known 
physical relationships) from model $\mathbf{m}$, as this is assumed to be a reasonable approximation to the pdf of uncertainties or errors in measured data.

\subsection{Mixture density networks}
Mixture density networks (MDNs) are a class of neural networks which combine a mixture model and a feed-forward network, so that the network can be trained to output probability density functions for any given input. A mixture model is represented as a linear combination of kernel functions, and can be used to provide accurate approximations to a general distribution $p(\mathbf{m}|\mathbf{d})$,
	\begin{equation}
		p(\mathbf{m}|\mathbf{d}) \approx \sum_{i=1}^{N}\alpha_{i}(\mathbf{d})\phi_{i}(\mathbf{m}|\mathbf{d})
		\label{eq:mixture}
	\end{equation}
where $\phi_{i}(\mathbf{m}|\mathbf{d})$ represents the $i^{th}$ kernel function, $\alpha_{i}$ is called the mixture coefficient which describes the relative importance of each kernel and satisfy $\sum_{i=1}^{N}\alpha_{i}=1$, and $N$ is the number of components in the mixture \citep{mclachlan1988mixture}. Various kernel functions can be used in practice. In this study we choose a Gaussian kernel function
	\begin{equation}
		\phi_{i}(\mathbf{m}|\mathbf{d}) = \frac{1}{\prod_{k=1}^{c}\left( \sqrt{2\pi}\sigma_{ik}(\mathbf{d}) \right)}
		\mathrm{exp}\left\{ -\frac{1}{2} \sum_{k=1}^{c} \frac{\left(m_{k} - \mu_{ik}(\mathbf{d}) \right)^{2}}{\sigma_{ik}^{2}(\mathbf{d})} \right\}
		\label{eq:gaussian_kernel}
	\end{equation}
where $c$ is the dimensionality of model $\mathbf{m}$, $\mu_{ik}$ represents the $k^{th}$ element of the mean of the $i^{th}$ Gaussian kernel, and $\sigma_{ik}$ is the corresponding standard deviation. Note that in equation (\ref{eq:gaussian_kernel}) we have neglected correlation between parameters of model $\mathbf{m}$. In principle, however, this assumption does not affect the accuracy of the approximation because this kind of mixture models are sufficient to approximate any density function to arbitrary accuracy, given that the mixture coefficients, the mean and variance of Gaussian kernels are properly selected \citep{mclachlan1988mixture, bishop2006pattern}. 

With above definition one can train a neural network which takes data $\mathbf{d}$ as input and outputs an approximation of the pdf $p(\mathbf{m}|\mathbf{d})$ parameterized by the mixture coefficient $\alpha_{i}$, the mean $\mu_{ik}$ and the standard deviation $\sigma_{ik}$, by using a set of data-model pairs {$(\mathbf{d}_{i},\mathbf{m}_{i})$} where $\mathbf{d}_{i}$ is calculated from model $\mathbf{m}_{i}$ using a known forward function. The network training is performed by minimizing the negative log likelihood of the pdf in equation (\ref{eq:mixture}), which is equivalent to maximizing the likelihood of the pdf. Once trained, the network can be used to predict the posterior pdf $p(\mathbf{m}|\mathbf{d}_{obs})$ for any observed data $\mathbf{d}_{obs}$. For a more detailed description of MDNs, we refer the reader to \cite{bishop2006pattern}, or to \cite{meier2007aglobal} with applications in geophysics.

\subsection{Graph mixture density networks}
In principle, any neural networks can be used in MDNs. For example, multi-layer perceptrons (MLP) and convolutional neural networks (CNNs) have been used in MDNs to solve inverse problems in geophysics \citep{meier2007aglobal, shahraeeni2011fast, earp2020probabilistic}. However, these networks usually require fixed dimensions of inputs, and hence can only be applied to problems with fixed number of data, or to those whose data can be interpolated to a certain number of dimensions. In reality, many problems have variable sizes of data, for example, in seismic tomography the number of travel time data can vary from time to time because of noise or malfunction of instruments. In order to extend applications of MDNs in such cases, we introduce graph neural networks (GNNs) to enable variable sizes of inputs.

\begin{figure}
\includegraphics[width=.9\linewidth]{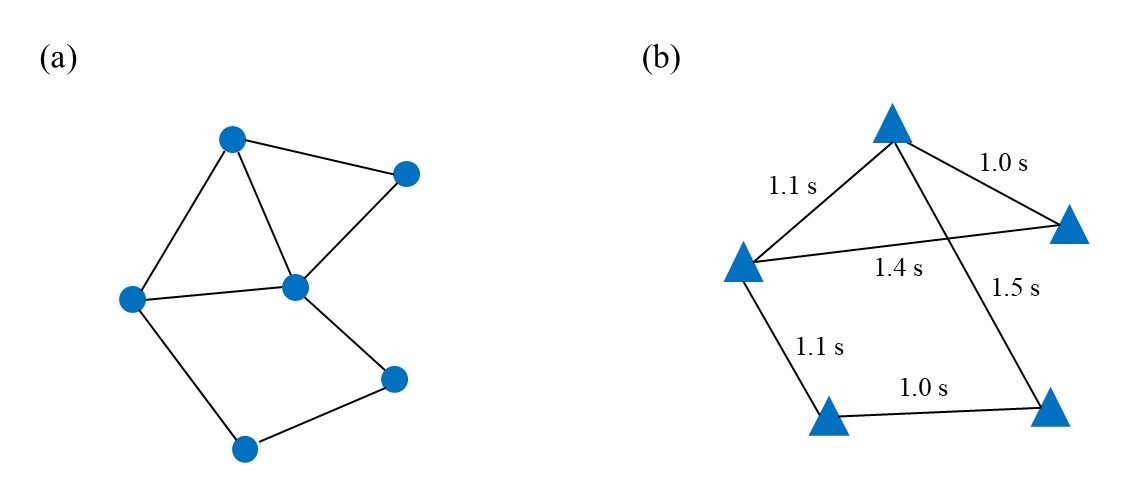}
\caption{(\textbf{a}) Illustration of a graph, where solid circles represent nodes and black lines represent edges between nodes. (\textbf{b}) Graph representation of travel time data in seismic surface wave tomography, where stations are treated as nodes and travel times between stations are treated as edge features.}
\label{fig:graph_illustration}
\end{figure}

Graphs are a type of data structure which describes a set of objects (nodes) and their relationships (edges), where edges can be directed or undirected (Figure \ref{fig:graph_illustration}a). Graphs provide flexible representations to many complex systems, for example social networks in social science, molecules in chemistry, protein in biology and systems in many other fields. In this study, we use graphs to represent travel time data in seismic surface wave tomography, where stations are denoted as nodes and undirected edges represent existence of travel time data between station pairs (Figure \ref{fig:graph_illustration}b). The coordinates of each station are set as node features and the travel time data between station pairs are denoted as edge features. In this way the graph provides a flexible representation of travel time data since a graph can have variable sizes of nodes (stations), and nodes can have different number of neighbors, that is, different number of travel time data.  

GNNs are neural networks that manipulates graphs. A variety of different types of GNNs have been proposed, for example, spectral networks \citep{bruna2013spectral}, graph convolutional networks \citep{kipf2016semi}, graph attention networks \citep{velivckovic2017graph}, graph transformers \citep{shi2020masked}, among many others \citep{scarselli2008graph, wu2020comprehensive}. In this study we utilize graph transformers to design GNNs \citep{shi2020masked}. Assume $\mathbf{x}_{i}$ as the $i^{th}$ node feature, and $\mathbf{e}_{ij}$ as the edge feature between the $i^{th}$ and $j^{th}$ node, a graph transform layer can be defined as:
	\begin{equation}
        \mathbf{x}^{'}_{i} = \mathbf{W}_{1}\mathbf{x}_{i} + \sum_{j\in\mathcal{N}(i)} \alpha_{ij}(\mathbf{W}_{2}\mathbf{x}_{j} + \mathbf{W}_{5}\mathbf{e}_{ij})
		\label{eq:graph_transform}
	\end{equation}
where $\mathbf{x}^{'}_{i}$ is the output of the transformer layer, $\mathcal{N}(i)$ represents neighbors of the $i^{th}$ node, that is, nodes that are connected with the $i^{th}$ node, and $\alpha_{ij}$ are called attention coefficients which can be computed using:
	\begin{equation}
		\alpha_{ij} = \mathrm{softmax}\left( \frac{(\mathbf{W}_{3}\mathbf{x}_{i})^{\mathrm{T}}(\mathbf{W}_{4}\mathbf{x}_{j}+\mathbf{W}_{5}\mathbf{e}_{ij})}{\sqrt{d}} \right)
		\label{eq:attention_coefficients}
	\end{equation}
where $d$ is the size of $\mathbf{W}_{3}\mathbf{x}_{i}$, and $\{\mathbf{W}_{m},m=1,2,3,4,5\}$ are tensors that represent trainable weights in the neural network. This transformer operator exploits both node features and edge features, and is therefore suitable for seismic tomography where edge features (travel times) are essential inputs. The above equations can be easily extended to multi-head attentions to improve expressiveness of the transformer layer,
	\begin{equation}
		\begin{aligned}
		\mathbf{x}^{'}_{i} &=& \bigg\Vert_{c=1}^{C}\bigg[\mathbf{W}_{1}^{c}\mathbf{x}_{i} + \sum_{j\in\mathcal{N}(i)} \alpha_{ij}^{c}(\mathbf{W}_{2}^{c}\mathbf{x}_{j} + \mathbf{W}_{5}^{c}\mathbf{e}_{ij})\bigg] \\
		\alpha_{ij}^{c} &=& \mathrm{softmax}\left( \frac{(\mathbf{W}_{3}^{c}\mathbf{x}_{i})^{\mathrm{T}}(\mathbf{W}_{4}^{c}\mathbf{x}_{j}+\mathbf{W}_{5}^{c}\mathbf{e}_{ij})}{\sqrt{d}} \right)
		\end{aligned}
		\label{eq:graph_multhead_transform}
	\end{equation}
where $\Vert$ represents concatenation operator of $C$ vectors (head attentions), and the superscript $c$ denotes the $c^{th}$ attention.

\begin{figure}
	\includegraphics[width=1.\linewidth]{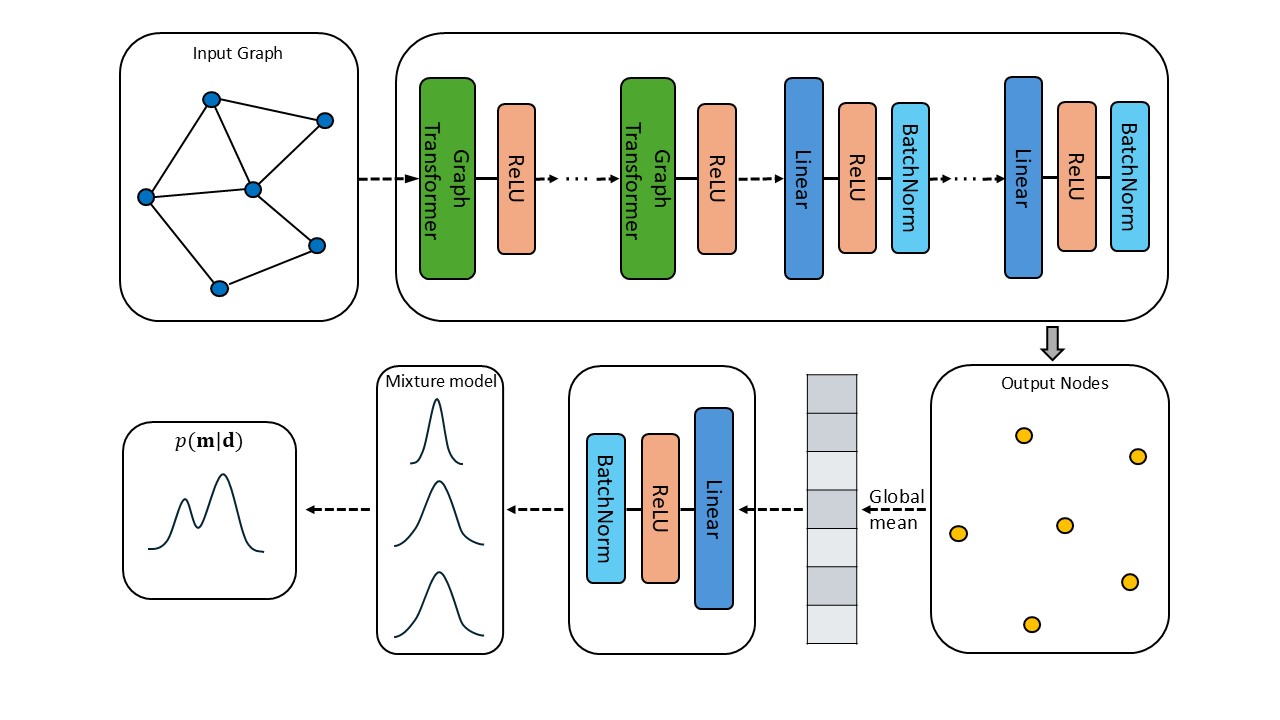}
	\caption{The architecture of the deep learning model designed for seismic tomography, which combines graph neural networks and mixture density networks. The model uses graph transformer layers and linear layers to update node features of the input graph, and take the mean of updated node features to predict the probability $p(\mathbf{m}|\mathbf{d})$ using a mixture model.}
	\label{fig:graphMDN_illustration}
\end{figure} 

GNNs and MDNs can be combined to create graph mixture density networks, which can predict conditional probabilities for graph inputs \citep{errica2021graph}. In this study we use graph MDNs to predict probability $p(\mathbf{m}|\mathbf{d})$ for travel time data $\mathbf{d}$ that are represented by a graph (Figure \ref{fig:graph_illustration}b). Figure \ref{fig:graphMDN_illustration} shows the designed deep learning model. The model uses graph transformer layers and graph linear layers to transform an input graph to an updated graph whose node features are calculated using both node and edge features of the input graph. These node features are then used to calculate the mean feature across all nodes, which are fed into a linear layer to output a mixture model that approximates the pdf $p(\mathbf{m}|\mathbf{d})$ (Figure \ref{fig:graphMDN_illustration}). We use ReLU activate functions for each graph transformer layer or linear layer. In addition, batch normalization is used after each linear layer. By using graph mixture density networks, this deep learning model can provide estimate of the probability $p(\mathbf{m}|\mathbf{d})$ for  data $\mathbf{d}$ with variable sizes.
 

\section{Synthetic tests}
\begin{figure}
\includegraphics[width=1.\linewidth]{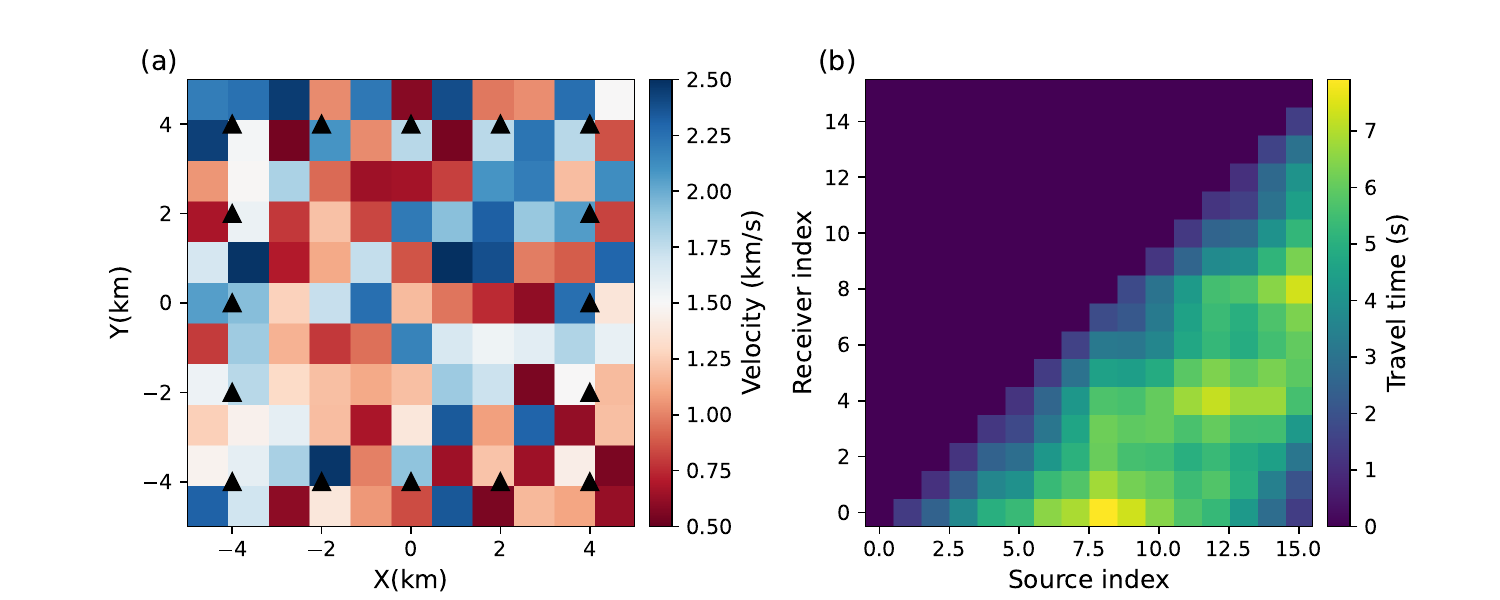}
\caption{(\textbf{a}) Receiver locations (black triangles) used in the synthetic travel time tomography experiment and an example velocity structure from the training set parameterized using a 9 $\times$ 9 grid. (\textbf{b}) The corresponding travel time field.}
\label{fig:synthetic_acquisition}
\end{figure}

To demonstrate the above method, we first train a graph MDN for a 2D travel time tomographic problem similar to those described in \cite{earp2020probabilistic} and \cite{zhang2021bayesianb} so that the results can be compared with their findings. The acquisition system contains 16 evenly distributed receivers in a shape of a square (Figure \ref{fig:synthetic_acquisition}a), each of which also acts as a virtual source to simulate a typical ambient noise interferometry experiment \citep{shapiro2005high}. The velocity model is parameterized using a 9 $\times$ 9 regular grid. At each grid point the prior distribution of the velocity is set as a Uniform distribution between 0.5 and 2.5 km/s. From this prior distribution we generate 200,000 velocity structures (Figure \ref{fig:synthetic_acquisition}a), and calculate the inter-receiver travel time data (Figure \ref{fig:synthetic_acquisition}b) for each structure using a fast marching method \citep{rawlinson2004multiple}. Gaussian noise with a standard deviation of 0.05 s is added to the data. Similarly to that in \cite{zhang2021bayesianb}, we add a halo of cells with random velocity around the receiver array, which will not be imaged but allows waves to travel both inside and outside the array. The travel time data for each velocity structure is represented as a graph as described in above section. To make edge feature values within a smaller range, we use the average velocity calculated from the travel time and inter-receiver distance as the edge feature for each receiver pair. For network training we use 90 percent of those velocity structure and data pairs as training data and the remaining 10 percent as test data for independent evaluation of network performance.
 
We use a graph MDN similar to that illustrated in Figure \ref{fig:graphMDN_illustration}, which consists of four graph transformer layers, four graph linear layers and a mixture model with 10 Gaussian kernels (details in Appendix A1). The training is performed using ADAM optimizer \citep{kingma2014adam} for 500 iterations with a learning rate of 0.0001 and a batch size of 1,000. To enable variable input sizes, we randomly drop stations pairs (edges) with a rate of 0.1 at each training iteration. The trained neural network is then used to predict posterior pdf $p(\mathbf{m}|\mathbf{d}_{\mathrm{obs}})$ for any observed data $\mathbf{d}_{\mathrm{obs}}$. To better understand performance of the method, we compare the results with those obtained using McMC. For any specific data $\mathbf{d}_{\mathrm{obs}}$ we use a standard adaptive Metropolis-Hastings algorithm \citep{haario2001adaptive, salvatier2016probabilistic} with a total of 6 chains, each of which contains 1,000,000 samples including a burn-in period of 500,000; the burn-in samples are discarded and every 10th of the remaining 500,000 samples are used to calculate statistics (mean and standard deviations) of the posterior pdf. 

\begin{figure}
\includegraphics[width=1.\linewidth]{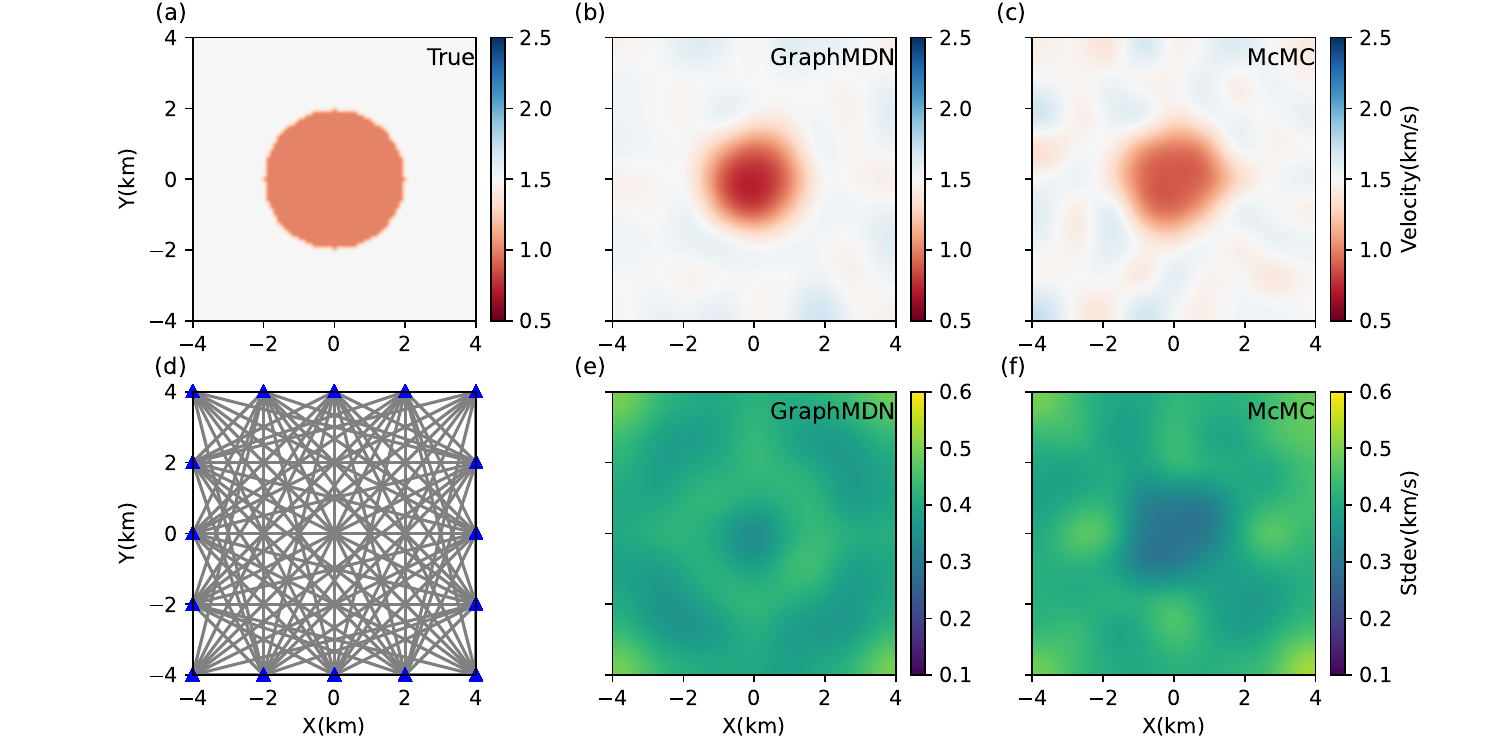}
\caption{(\textbf{a}) The true velocity structure. (\textbf{b}) and (\textbf{c}) show the mean velocity structure obtained using graph MDN and McMC respectively. (\textbf{e}) and (\textbf{f}) show the corresponding standard deviation. (\textbf{d}) shows straight ray paths (station pairs) used in the inversion.}
\label{fig:circle_full}
\end{figure}

\begin{figure}
	\includegraphics[width=1.\linewidth]{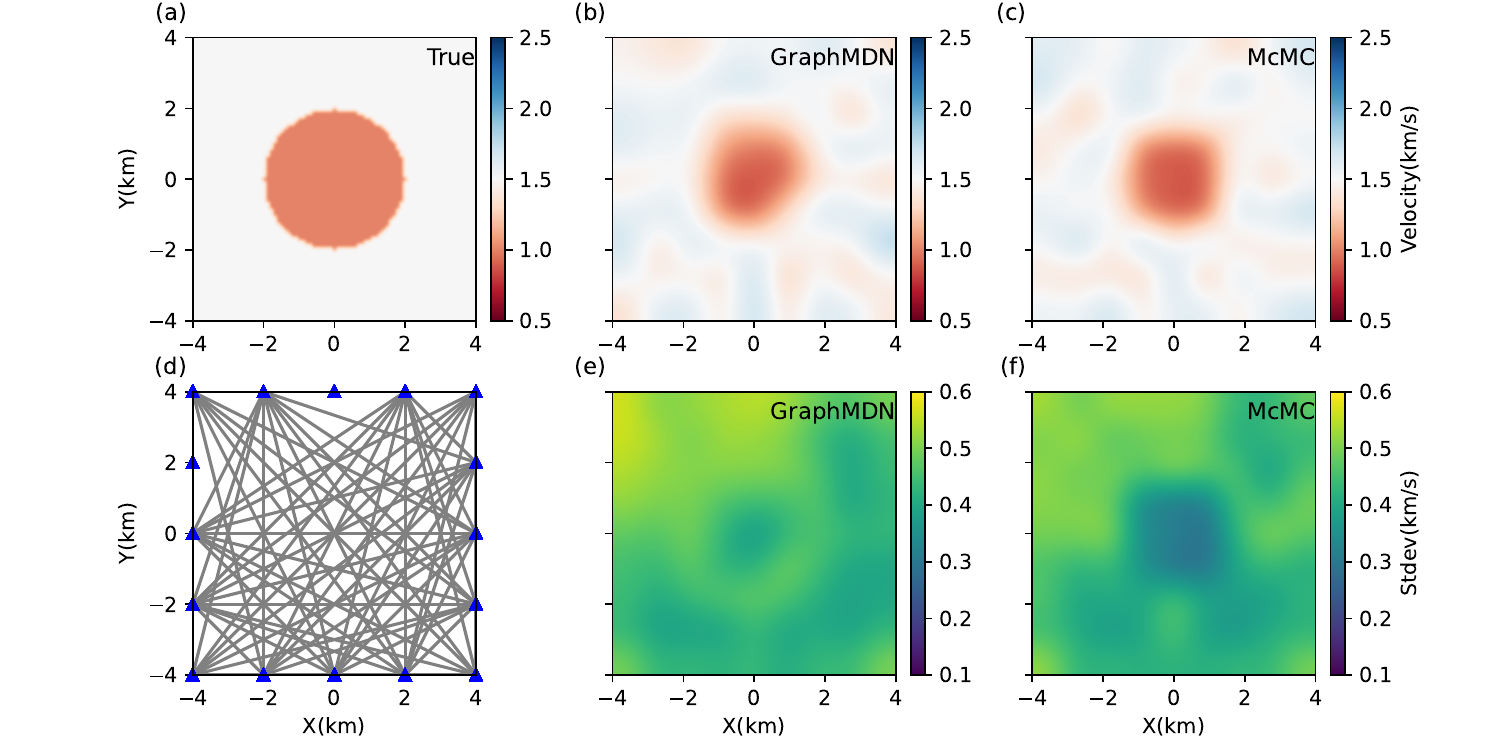}
	\caption{Results obtained in a similar experiment as in Figure \ref{fig:circle_full} but using part of the data as shown in (d). Keys as in Figure \ref{fig:circle_full}.}
	\label{fig:circle_part}
\end{figure}

As a first example, we study the posterior pdf for the data generated using a smooth velocity structure similar to that in \cite{earp2020probabilistic} and \cite{zhang2021bayesianb}. The velocity structure is deliberately designed to have a low velocity anomaly in the center within a homogeneous background (Figure \ref{fig:circle_full}a), such that the structure contains both perfectly smooth regions and a sharp, spatially coherent boundary. Note that the smooth regions are not represented by any structures in the training set (see an example in Figure \ref{fig:synthetic_acquisition}a), so any reasonable solution must be interpolated between training examples. The travel time data are computed using a higher resolution grid (81 $\times$ 81), and are fed into trained neural networks to predict the posterior pdf, and are also used in McMC to generate posterior samples. To analyze performance of the neural network in the case of variable input sizes, we also created a new data set by randomly dropping part of the station pairs (Figure \ref{fig:circle_part}d), and inverted the data set for the posterior pdf using both methods. For graph MDN the new data set is directly fed into the trained neural network to predict the posterior pdf, whereas for McMC the inversion is re-conducted in the same way as described above.  

Figure \ref{fig:circle_full} shows results obtained using the full data set. For ease of visual interpretation, we constructed smooth tomographic images by using interpolation. Overall the two methods produce similar mean (Figure \ref{fig:circle_full}b and \ref{fig:circle_full}c) and standard deviation structures (Figure \ref{fig:circle_full}e and \ref{fig:circle_full}f). For example, both mean structures show the circle low velocity anomaly in the center, and small velocity variations around the anomaly which may be caused by lower sensitivity in these regions or by data noise. The standard deviation structures show lower uncertainties in the center, and higher uncertainties around the low velocity anomaly which have been observed previously and reflect uncertainty of the shape of the anomaly \citep{galetti2015uncertainty, zhang2020seismic}. There are also higher uncertainties in the four corners of the grid which are caused by relatively poor data coverage (Figure \ref{fig:circle_full}d).

Figure \ref{fig:circle_part} shows results obtained using part of the data set. Similarly as above, the two methods produce similar mean and standard deviation structures. For example, in the northwestern region both results show higher uncertainties because of poor ray coverage (Figure \ref{fig:circle_part}d), whereas in other areas the results show similar features as those obtained using the full data set (Figure \ref{fig:circle_full}). However, the magnitude of uncertainty becomes higher than that obtained using the full data set across the whole region, which reflects the effects caused by lower number of data in the inversion. These results therefore demonstrate that graph MDNs can be trained to predict probability distributions for variable sizes of inputs.

There are also features in the results of McMC which cannot be clearly observed in the results of graph MDN. For example, four high standard deviation anomalies can be observed above, below, left and right of the central anomaly (Figure \ref{fig:circle_full}f and \ref{fig:circle_part}f), which are not clearly visible in the results of graph MDN (Figure \ref{fig:circle_full}e and \ref{fig:circle_part}e). Note that this is a high dimensional (81) problem, and consequently a large number of samples are required to adequately sample the parameter space. It is therefore possible that neither the training set nor the McMC samples is sufficient to explore the significant areas of the tomographic solution due to the curse of dimensionality \citep{curtis2001prior}. As a result, the two solutions can have different detailed structures. Nevertheless, the broad characters of the mean and standard deviation structures match those found previously \citep{earp2020probabilistic, zhang2021bayesianb}, so at least these statistics are (approximately) correct.

\begin{figure}
	\includegraphics[width=1.\linewidth]{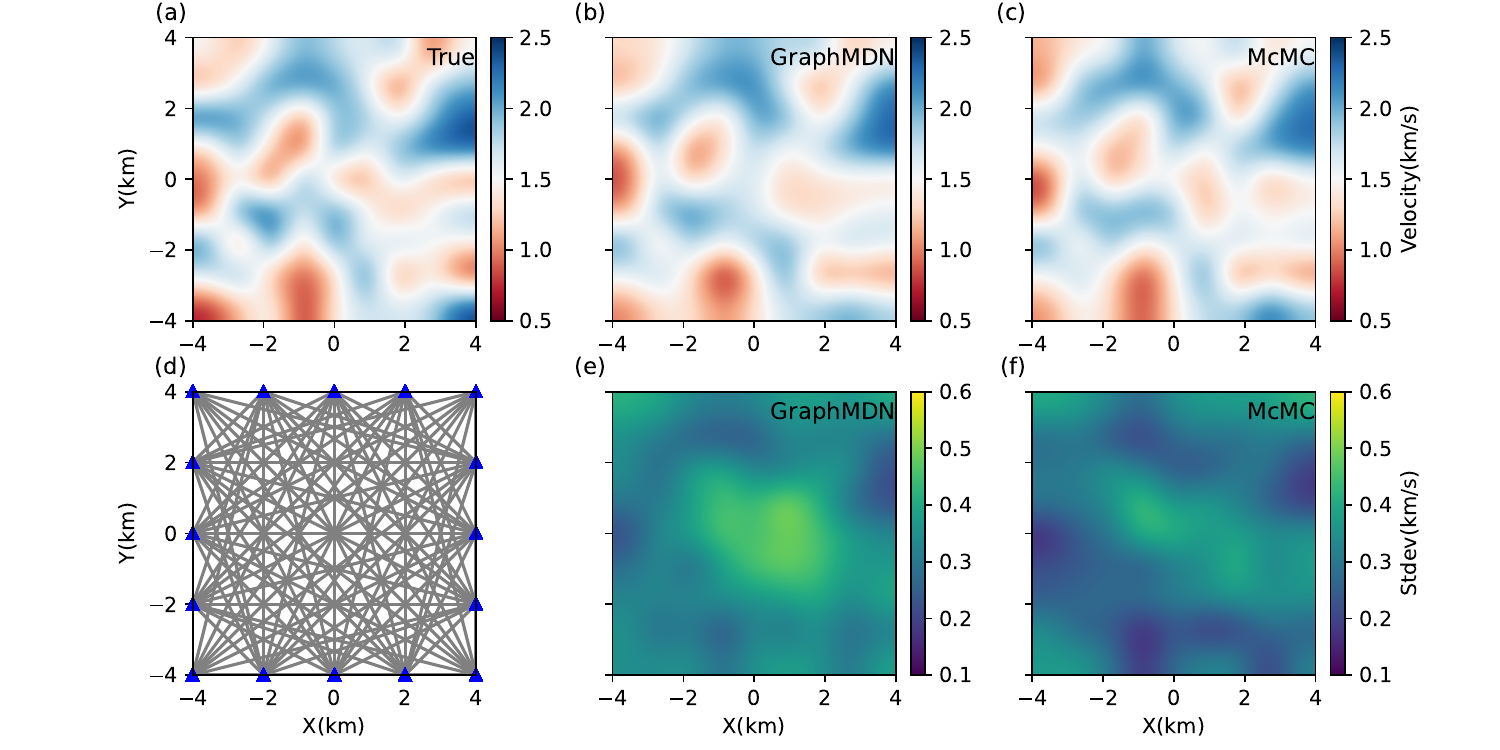}
	\caption{Results for a random heterogeneous velocity structure obtained using the full data set. Keys as in Figure \ref{fig:circle_full}.}
	\label{fig:random_full}
\end{figure}

\begin{figure}
	\includegraphics[width=1.\linewidth]{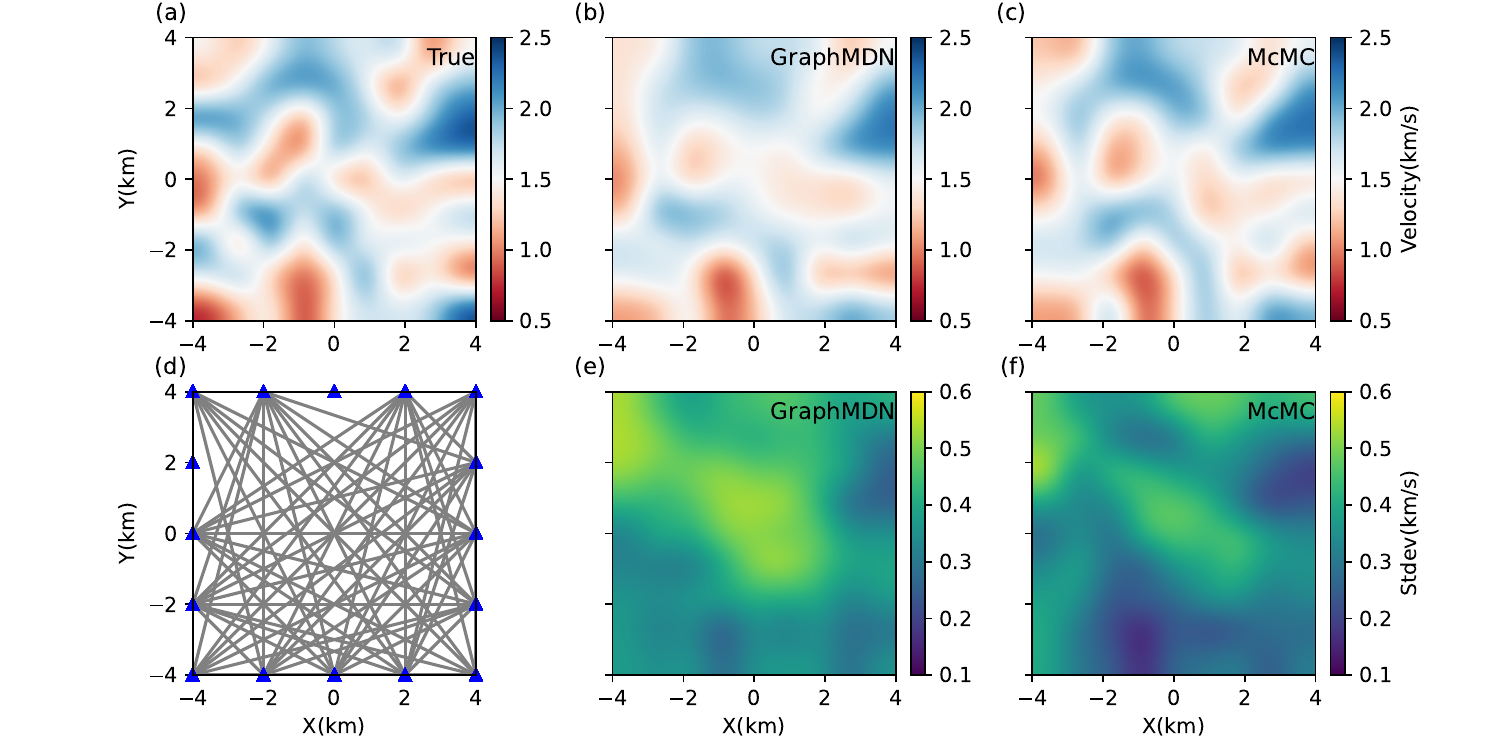}
	\caption{Results for a random heterogeneous velocity structure obtained using part of the data set. Keys as in Figure \ref{fig:circle_full}.}
	\label{fig:random_part}
\end{figure}

To explore generalization properties of the trained neural network, we conducted another test using data generated from a random velocity structure (Figure \ref{fig:random_full}a) similar to that in \cite{zhang2021bayesianb}. The travel time data are calculated using the fast marching method, and a new data set with different ray path distribution is created from these data by keeping the same station pairs as in Figure \ref{fig:circle_part}d. Those data sets are then fed into the trained neural network to produce estimates of the posterior pdf, and are also inverted using McMC in the same way as described above.

Figure \ref{fig:random_full} shows the results obtained using the full data set. Overall the two methods produce similar mean and standard deviation structures. Both mean structures are largely the same as the true structure. Both standard deviation maps show lower uncertainties at locations of the low velocity anomalies in the west and south side, and at locations of the high velocity anomalies in the northeast and north of the region. By contrast, in the center and four corners the results show higher uncertainties which reflects low data sensitivities in these areas.

Figure \ref{fig:random_part} shows the results obtained using part of the data set. Similarly as above, the two methods produce largely consistent mean and standard deviation structures. For example, both results show less velocity variation and higher uncertainties in the northwest because of lower ray coverage (Figure \ref{fig:random_part}d). In addition, as observed in the first example the magnitude of uncertainty is higher across the whole region compared to that obtained using the full data set because of the lower number of data. Note that due to insufficient training data or possible lack of convergence of McMC the standard deviations show some detailed features that differ between the two methods. For example, the low uncertainty anomaly in the north side of the McMC results cannot be clearly observed in the standard deviation maps of the graph MDN. Nevertheless, the results are largely consistent, which again demonstrates the ability of graph MDNs to take variable sizes of inputs and predict accurate posterior estimates.

Note that graph MDNs obtained above results using far less computational cost than McMC. For graph MDNs, it took about 0.24 hours to generate the training set using 6 Intel(R) Xeon(R) Gold 6258R CPU cores, and about 7 hours to train the neural network on a NVIDIA GeForce RTX 4090 GPU card. Once trained, the network can provide very efficient estimate of the posterior distributions. For example, the above trained network took 0.6 s to predict a posterior estimate for one specific data set. By contrast, for each data set McMC took about 21 hours to produce the above results using the same set of CPU cores. Thus trained graph MDNs provide an important tool in scenarios where many repeated inversions with variable sizes of data are required to predict posterior distributions of the subsurface variations over space \citep{meier2007aglobal} or time \citep{cao2020near}.
     
\begin{figure}
	\includegraphics[width=1.\linewidth]{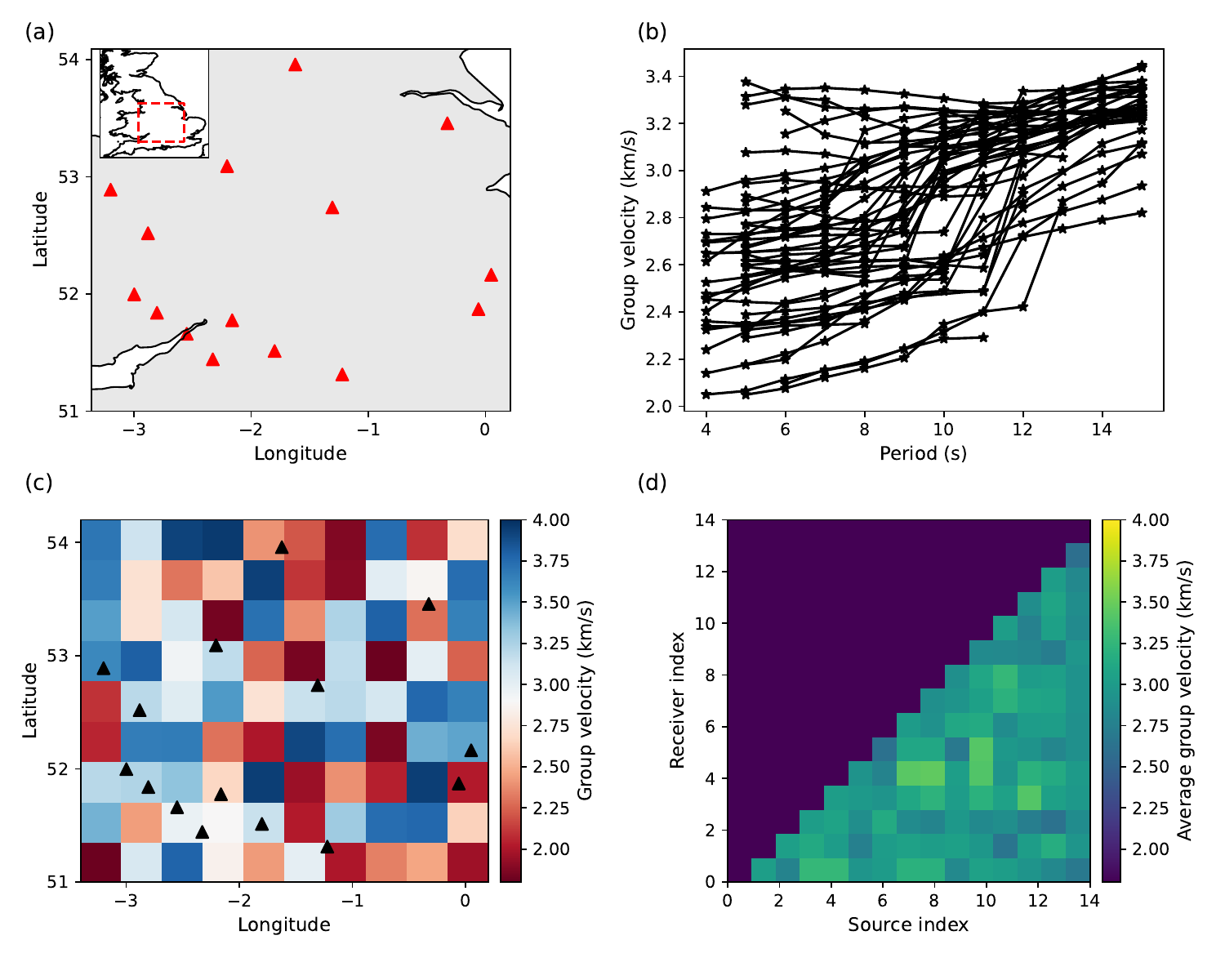}
	\caption{(\textbf{a}) Receivers (red triangles) used in this study. The inset map denotes the location of the study area. (\textbf{b}) Group velocity dispersion curves extracted from ambient noise data between each station pair, which are computed using station pair distances and travel times at different periods. The number of data varies between different periods. (\textbf{c}) An example velocity structure from the training set and (\textbf{d}) the corresponding group velocity (travel times) between each station pair.}
	\label{fig:uk_setting}
\end{figure}

\section{Real data application}
\begin{figure}
	\includegraphics[width=1.\linewidth]{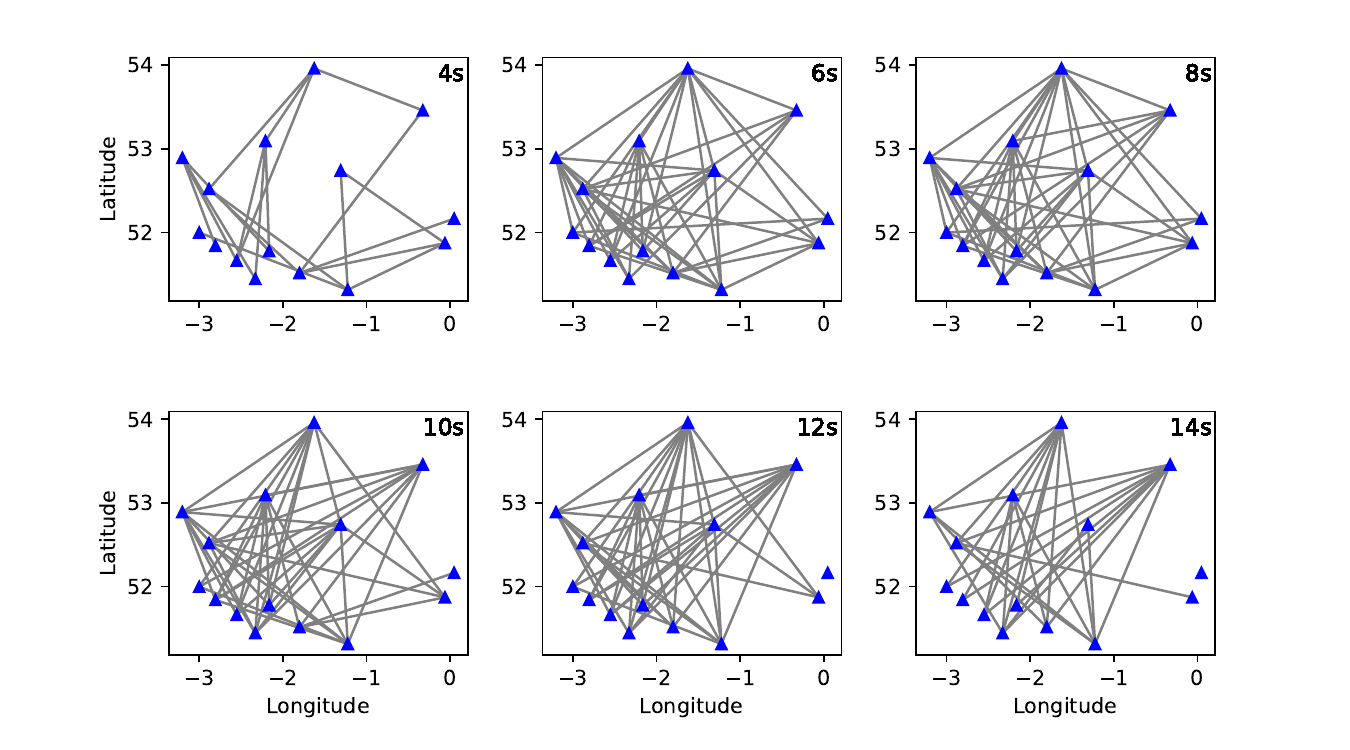}
	\caption{Ray paths used for tomography at different periods.}
	\label{fig:uk_raypath}
\end{figure}

In this section we train a graph MDN to predict posterior distributions of group velocity at different periods for Love waves extracted from ambient noise data of South England (Figure \ref{fig:uk_setting}a), which has complex geological structures compared to the rest of the British Isles \citep{nicolson2012seismic,nicolson2014rayleigh,galetti2017transdimensional}. We use ambient noise data recorded with 15 stations (Figure \ref{fig:uk_setting}a) in 2006 - 2007 and in 2010. To obtain Love waves, the horizontal components (N and E) were first rotated to the transverse and radial directions, and the obtained transverse data were cross correlated to produce Love waves for all possible station pairs. To improve the signal-to-noise ratio, cross-correlations were calculated using daily data and were linearly stacked over the total recording period. We then estimated travel time data of group velocity at different periods from those love waves using the multiple-filter analysis method \citep{herrmann2013computer}, which are represented as path-averaged group velocities when assuming a great circle ray path for each station pair (Figure \ref{fig:uk_setting}b). The uncertainty of those travel times are estimated from daily cross correlations. In order to improve quality of the data, we only kept measurements whose uncertainties are smaller than 5 percent of the total travel time. In addition, due to the far-field approximation that is implicit in the ambient noise interferometry method, group velocity measurements between station pairs whose distances are less than 3 wavelengths were discarded \citep{bensen2007processing}. More details of the data processing procedure can be found in \cite{galetti2017transdimensional}. Note that due to the above quality control procedure, the number of estimated travel times usually varies with periods (Figure \ref{fig:uk_raypath}). Thus, a deep learning model that can take inputs with various sizes is required in order to predict posterior distributions of group velocity for different periods.

Similarly to that in \cite{zhang2021bayesianb}, we parameterize group velocity models using a 9 $\times$ 10 regular grid of cells with a spacing of $0.4^\circ$ in both longitude and latitude directions (Figure \ref{fig:uk_setting}c). The prior pdf for group velocity is set to be a Uniform distribution between 1.8 and 4.0 km/s according to the path-averaged group velocities (Figure \ref{fig:uk_setting}b) and to the previous research of \cite{galetti2017transdimensional}. For the likelihood function we use a Gaussian distribution with a diagonal covariance matrix to represent the data noise. The standard deviation of the Gaussian distribution is computed by assuming a consistent noise to travel time ratio across periods for each station pair, which is calculated as the average ratio across different periods from uncertainty estimated using independent travel time measurements of daily cross correlations \citep{galetti2017transdimensional}.

As in the synthetic tests, we generate 200,000 velocity models from the prior pdf (Figure \ref{fig:uk_setting}c), and calculate inter-station travel times using the fast marching method for each velocity model (Figure \ref{fig:uk_setting}d). The corresponding noise of those data is generated from a Gaussian distribution that has zero mean and the same covariance matrix as the likelihood function. Again 90 percent of those data are used as training data and the remaining 10 percent as test data. For network design we use a similar graph MDN to that used in the above synthetic examples, which consists of five graph transformer layers, four graph linear layers and a mixture model with 10 Gaussian kernels (details in Appendix A2). The network is then trained using ADAM optimizer for 1,000 iterations with a learning rate of 0.0004 and a batch size of 500. During training we randomly drop station pairs with a rate of 0.6 at each iteration according to the number of travel times observed in real data. The trained network is thereafter used to predict posterior pdfs of group velocities for those observed data at different periods.

For comparison, we also solved the set of tomographic problems at different periods using McMC. Similarly to the previous examples, we run 6 independent chains, each of which contains 1,000,000 samples including a burn-in period of 500,000. The burn-in samples were discarded and every 10th of the remaining samples was collected as the final set of McMC samples. The mean and standard deviation are then calculated using those samples.

\begin{figure}
	\includegraphics[width=1.\linewidth]{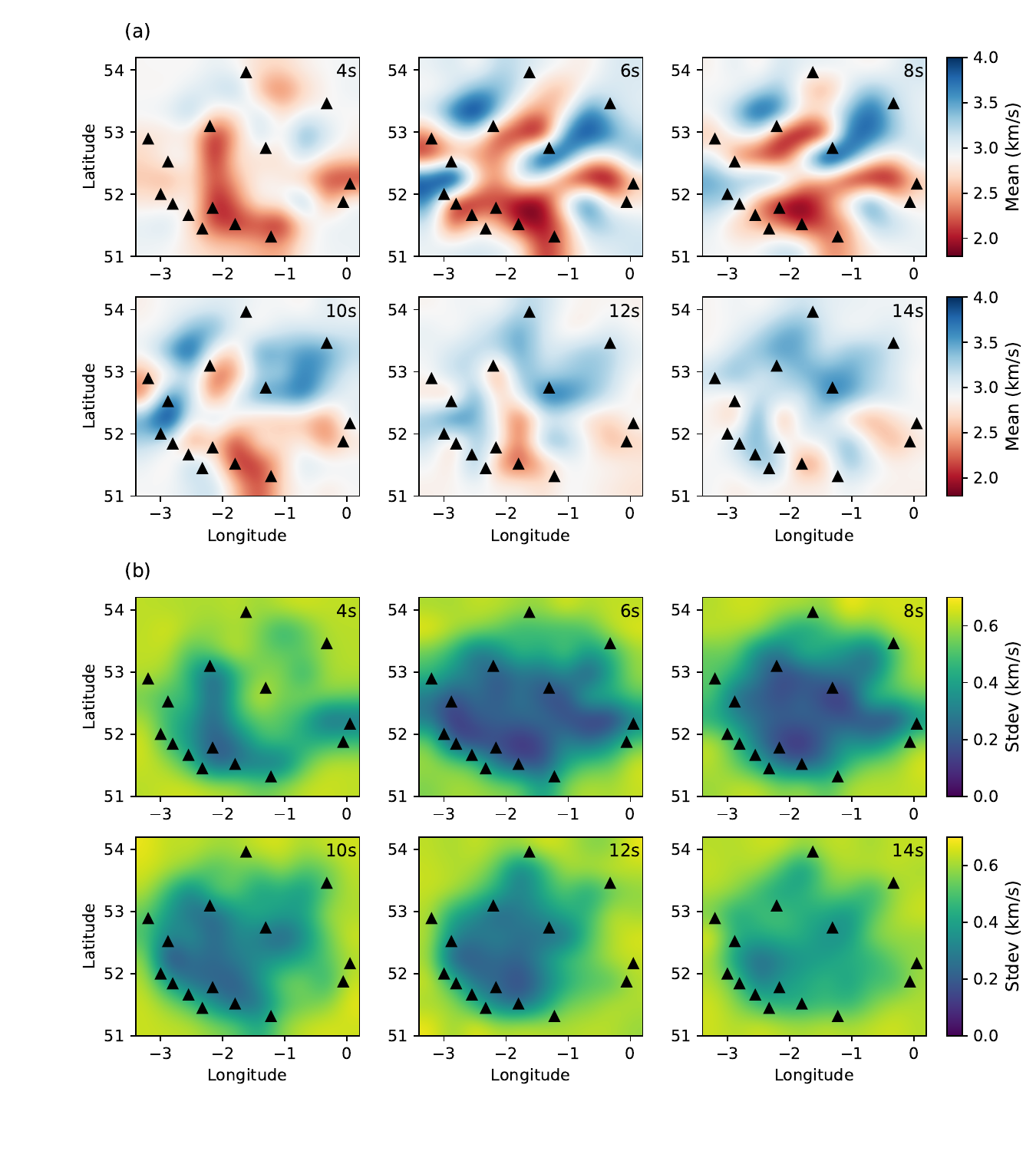}
	\caption{(\textbf{a}) The mean and (\textbf{b}) standard deviation maps obtained using graph MDN at periods of 4 s, 6 s, 8 s, 10 s, 12 s and 14 s.}
	\label{fig:uk_graphmdn}
\end{figure}

Figure \ref{fig:uk_graphmdn} shows the mean and standard deviation maps obtained using graph MDN at different periods. Overall group velocities become larger with increasing periods, which reflects the general tendency of seismic velocity to increase with depth. The structures of these mean velocity maps generally reflect various geology features at shallow depths in the region, with sedimentary rocks being shown as low velocity anomalies while igneous and metamorphic complexes are displayed as high velocities as described in \cite{galetti2017transdimensional}. For example, between 4 and 10 s there is a low velocity anomaly in the south of the region (around $-1.8^{\circ}$E, $51.8^{\circ}$N), which is associated with the Midland Platform and has also been observed in the results obtained using reversible-jump McMC \citep{galetti2017transdimensional}. The anomaly is still visible in the results at 12 and 14 s but with decreasing sizes as period increases, which may reflect the higher crustal thickness of the area compared to other regions in South England \citep{chadwick1998seismic, tomlinson2006analysis}. To the north of this anomaly, another low velocity anomaly  can also be observed between 4 and 10 s which is associated with the Cheshire Basin (around $-2^{\circ}$E, $53^{\circ}$N). The anomaly disappears at 12 and 14 s, which likely suggests a smaller thickness of the sedimentary basin. By contrast, the low velocity anomaly in the east associated with the Anglian Basin (around $-0.2^{\circ}$E, $52.1^{\circ}$N) is visible across all the periods.

In the northwest there is a high velocity anomaly across all the periods following an approximately northeast-southwest trend around $-2.5^{\circ}$E, $53.5^{\circ}$N, which probably reflects the limestones of the Pennines \citep{galetti2017transdimensional}. Note that at 4 s period the velocity value of the anomaly is lower than those at other periods. This is likely caused by low ray coverage at the region at 4 s (Figure \ref{fig:uk_raypath}), which is also reflected by higher uncertainties in the standard deviation maps (Figure \ref{fig:uk_graphmdn}). A high velocity anomaly with similar trend can also be observed in the northeast of the region (around $-1.5^{\circ}$E, $53^{\circ}$N), which has been found in several previous studies \citep{nicolson2014rayleigh, galetti2017transdimensional} and is related to the northern limit of the Anglo-Brabant Massif \citep{nicolson2014rayleigh}. As suggested in \cite{galetti2017transdimensional}, the feature may reflect the granitic batholiths and dykes underneath the area, or may be interpreted as the evidence of Proterozoic basement in an area of thin sedimentary cover. Note that due to the low ray coverage, the velocity value of the anomaly is lower at 4 s period.  

The standard deviation maps in Figure \ref{fig:uk_graphmdn}b reflect the uncertainty of those group velocity estimates. Overall the results show high uncertainties in areas outside the receiver array because few ray paths go through these regions. Across the set of periods, the uncertainties appear to be higher as periods increase. This is because group velocities are generally higher at longer periods, and consequently the standard deviations have large magnitudes given the same level of travel time uncertainty. However, at 4 s period the standard deviations are higher than those at 6 and 8 s period, which clearly reflects the low ray coverage at the period as discussed above. The uncertainties are lower at locations of the low velocity anomalies in the middle, south and east of the region at 4 s period which suggests that those anomalies are well constrained by the data (Figure \ref{fig:uk_raypath}), whereas at the location of the low velocity anomaly in the north the standard deviation map show higher uncertainties which is caused by low ray coverage in the area (Figure \ref{fig:uk_raypath}). At 6 and 8 s period the standard deviation maps show similar features. For example, both results show lower uncertainties in the southwest because of high ray coverage. In addition, at the location of the low velocity anomaly in the east and the location of the high velocity anomaly in the northeast, despite of their relatively low ray coverage (Figure \ref{fig:uk_raypath}) the results show lower uncertainties which suggests that these low and high velocity features are strongly constrained by the data. Similar low uncertainty features can be observed in most part of the region at 10 and 12 s, except that at the location of the low velocity anomaly in the east there is no clearly visible lower uncertainty. This may be caused by the relatively low ray coverage (Figure \ref{fig:uk_raypath}), or reflect that a strong low velocity anomaly is not required by the data which can also be deduced from the higher magnitude of the anomaly. At 14 s the standard deviation map shows less variation in space compared to those at other periods, which is consistent to the smoother velocity structure in the mean map. 

\begin{figure}
	\includegraphics[width=1.\linewidth]{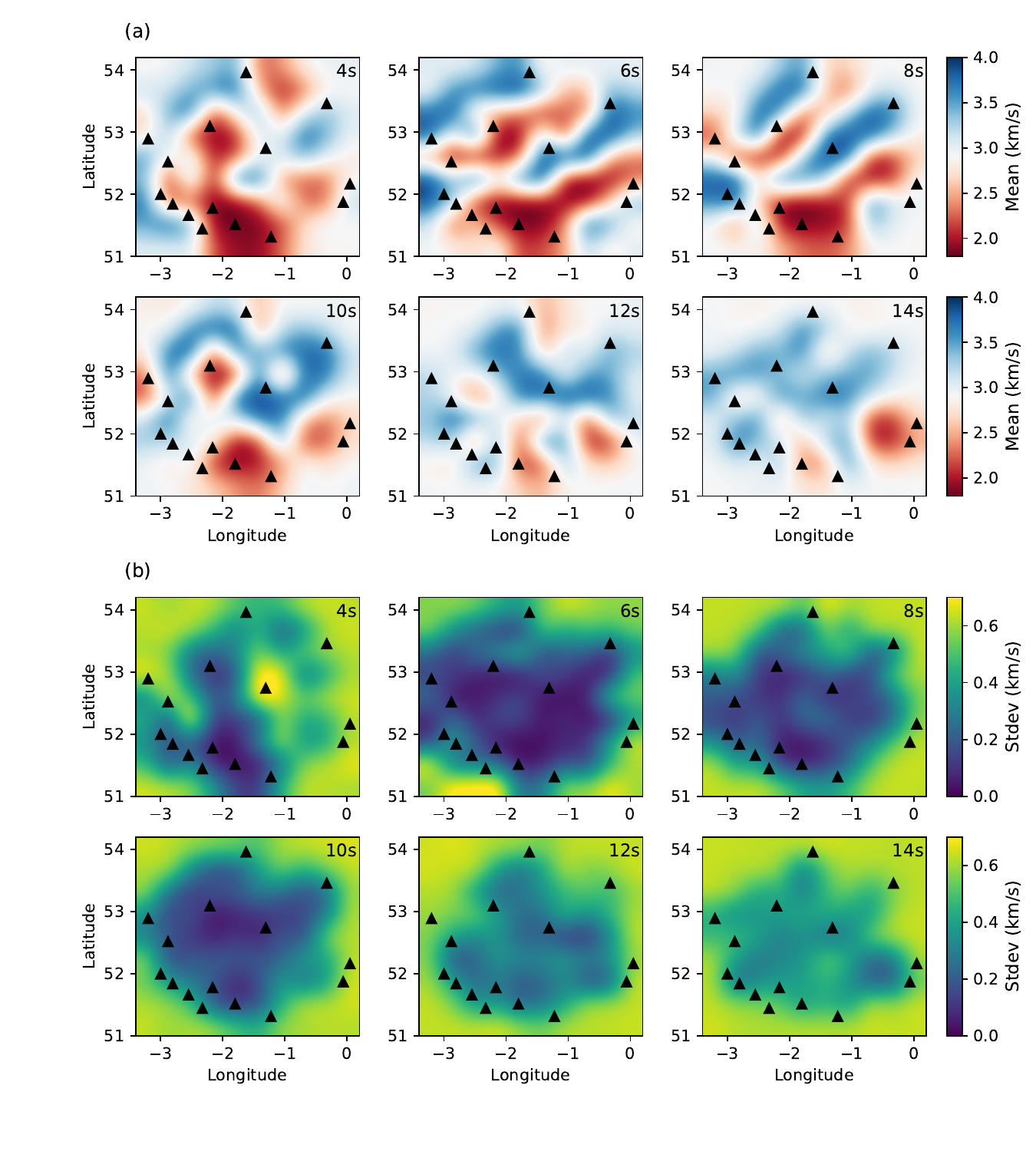}
	\caption{(\textbf{a}) The mean and (\textbf{b}) standard deviation maps obtained using McMC. Keys as in Figure \ref{fig:uk_graphmdn}.}
	\label{fig:uk_mcmc}
\end{figure}

Figure \ref{fig:uk_mcmc} shows the mean and standard deviation maps obtained using McMC. Overall the results show similar features to those obtained using graph MDNs. For example, the mean maps show similar low velocity anomalies in the south and east of the region across the set of periods, which are associated with the Midland Platform and the Anglian Basin respectively. The low velocity anomaly associated with the Cheshire Basin (around $-2^{\circ}$E, $53^{\circ}$N) can also be observed between 4 and 10 s and disappears at 12 and 14 s. In addition, there are also high velocity anomalies in the northwest and northeast of the region with an approximately northeast-southwest trend. However, note that there exists small details which are different in the two results. For example, magnitudes of velocities are slightly different at the periods of 4, 10, 12 and 14 s, which likely reflects the insufficient training of graph MDN.

Similarly to the results obtained using graph MDN, the standard deviation maps obtained using McMC show higher uncertainties at longer periods, and lower uncertainties at the locations of those low and high velocity anomalies. At 4 s period, the uncertainty is also higher because of the low ray coverage. However, the magnitude of uncertainty obtained using McMC is generally smaller than that obtained using graph MDN. This reflects the limitation of MDNs which become difficult to train in high dimensionality and consequently difficult to achieve high accuracy of approximation to the posterior pdf \citep{hjorth1999regularisation, cui2019multimodal, makansi2019overcoming, zhang2021bayesianb}. Nevertheless, the results obtained using graph MDNs are broadly consistent with those found using McMC.

Again graph MDNs required far less computational cost than McMC. For graph MDNs, it took about 0.15 hours to generate the training set using 6 Intel(R) Xeon(R) Gold 6258R CPU cores, and about 16.7 hours to train the neural network using a NVIDIA GeForce RTX 4090 GPU card. Once trained, the network predicts the posterior distribution for each data set in just 0.6 s. By contrast, McMC took about 16 hours to generate the above results for each data set using the same set of CPU cores, and thus 96 hours in total for all the periods used in the study. Note that graph MDNs will become more efficient if more periods are used, or if many inversions are required over different time periods.

\section{Discussion}
We demonstrated that graph MDNs can be trained to provide rapid solutions to seismic tomographic problems with variable sizes of data given a specific receiver geometry. However, the method can also be applied in scenarios where the number of stations is variable, for example, when adding more stations to an existing station array. In such cases one can train neural networks by adding a set of potential stations in the training data. From a theoretical point of view graph MDNs can also be trained using random station distributions, and the trained network may provide solutions for various tomographic problems at a specific site. In addition, it might also be possible that one can train a neural network to solve a certain type of tomographic problems at various sites as only the structure of station distributions, that is, relative locations between stations, is required. However, we note that this may require a very large network to be trained.

In the above study the results obtained using graph MDNs show higher uncertainty estimates than those obtained using McMC because of difficulties of training MDNs in high dimensionality \citep{hjorth1999regularisation, cui2019multimodal, zhang2021bayesianb}. To further improve accuracy, other networks such as invertible neural networks may be combined with graph networks to provide more accurate uncertainty estimates \citep{ardizzone2018analyzing, zhang2021bayesianb}. Alternatively, techniques such as variational auto-encoders \citep{kingma2013auto, laloy2017inversion} or generative adversarial networks \citep{goodfellow2014generative, mosser2020stochastic} may be used to reduce dimensionality of parameter spaces, or one can train graph MDNs to predict marginal distributions of only a few parameters as performed in \cite{earp2020probabilistic}.

In this study we used independent Uniform distributions for each cell which may become ineffective for large inverse problems \citep{curtis2001prior, de2017structure, earp2020probabilistic}. In practice where more knowledge about the subsurface is available, one can use a more informative prior distribution. For example, in scenarios where many inversions are required over time to detect subsurface changes, one can impose stronger prior constraints on model parameters \citep{zhang2024bayesian}. Smooth prior pdfs \citep{earp2020probabilistic} or structure-based prior pdfs \citep{de2017structure} may also be used to produce smoother velocity models. In addition, generative networks can be used to encode geological information into prior distributions \citep{laloy2017inversion, mosser2020stochastic}.

We used a fixed regular grid of cells to parameterize the velocity model which may become inappropriate if the trained network is applied to a different site. In addition, a fixed parameterization requires a graph-level operation, for example, the mean across different nodes used in this study, which may restrict expressiveness of the network. To increase flexibility, one can add more nodes to the graph and add velocity as node features in addition of location coordinates. By doing this, outputs of the network become to predict one of the node features (velocity). Consequently the network can provide a more flexible parameterization of the subsurface, and does not require a graph level operation. For data noise, we used a fixed noise to travel time ratio for different periods, which cannot take account of variable noise level for different data sets. To overcome this issue, one can add noise as an extra edge feature such that noise also becomes network inputs, similar to the way described in \cite{earp2020probabilistic}.

In this study we applied graph MDNs to 2D surface wave tomography, but it is straightforward to apply the method to other geophysical inverse problems. For example, the method can also be applied to body wave travel time tomographic problems by treating sources as graph nodes. We used graph transformers to implement graph neural networks, but other design of graph networks that incorporate both node and edge features may also be used, for example, message passing neural networks \citep{gilmer2017neural}, edge feature enhanced graph neural networks \citep{gong2019exploiting}, and node and edge graph neural networks \citep{yang2020nenn}.   

\section{Conclusion}
In this study we introduced graph MDNs to solve seismic tomographic problems. Graph MDNs combine graph neural networks with mixture density networks to provide estimate of posterior pdfs efficiently for graph data. We applied graph MDNs to seismic tomographic problems by representing travel time data as a graph such that the network can take variable sizes of data. We demonstrated the method using both synthetic and real data, and compared the results with those obtained using McMC. The results show that trained graph MDNs can provide accurate approximations of posterior pdfs obtained by McMC in seconds for variable distributions of data, such as posterior pdfs of group velocities at different periods where generally different number of data are available. We thus conclude that graph MDNs can be used to provide accurate and rapid solutions to seismic tomographic problems with variable sizes of data.

\begin{acknowledgments}
The authors thank the National Natural Science Foundation of China (42204055), the National Science and 
Technology Major Project (2024ZD1000403), and the Fundamental Research 
Funds for the Central Universities of CUGB for supporting this research.
\end{acknowledgments}

\bibliographystyle{gji}
\bibliography{graph.bib}

\appendix
\section{Network configuration}

\subsection{Network configuration for the synthetic test}

The graph MDN consists of four graph transformer layers, four graph linear layers and a mixture model with 10 Gaussian kernels as illustrated in Figure \ref{fig:graphMDN_illustration}. The number of channels of the 4 graph transformer layers is 32, 64, 128 and 256 respectively, and the number of head attentions is set to 1 for all layers. For the four graph linear layers, the number of hidden units is set to 500, 800, 600 and 1000 respectively. A global mean pooling layer is used to calculate the mean of updated node features, which are then fed into a linear layer with 1500 hidden units. 

\subsection{Network configuration for the real data example}
The graph MDN consists of five graph transformer layers, four graph linear layers and a mixture model with 10 Gaussian kernels. The number of channels of the 5 transformer layers is set to 32, 64, 128, 256 and 512 respectively, and the number of head attentions is set to 2 for all layers. The number of hidden units of 4 graph linear layers is set to 800, 800, 1000 and 1000 respectively. For the linear layer after the global mean pooling, the number of hidden units is set to 1500. 

\label{lastpage}

\end{document}